\documentclass[preprint,12pt, numbers, compress]{elsarticle}

\usepackage[colorlinks=true, urlcolor=blue, anchorcolor=blue, citecolor=blue,filecolor=blue,linkcolor=blue,menucolor=blue
]{hyperref}

\usepackage{color}

\title{Collective behavior of independent scaled Brownian particles with renewal resetting} 

\usepackage{amssymb}
\usepackage{amsmath}

\journal{Physica A}

\begin{document}

\begin{frontmatter}

\author[inst1]{Ohad Vilk}
\ead{ohad.vilk@mail.huji.ac.il}

\author[inst1]{Baruch Meerson}
\ead{meerson@mail.huji.ac.il}

\affiliation[inst1]{
  organization={Racah Institute of Physics, Hebrew University of Jerusalem},
  city={Jerusalem},
  postcode={91904},
  country={Israel}
}

\begin{abstract}
We study fluctuations of an ensemble of $N$ independent particles undergoing anomalous diffusion with random renewal resetting.  The anomalous diffusion is modeled by the scaled Brownian motion (sBm): a Gaussian process, characterized by a power-law time dependence of the diffusion coefficient, $D(t)\sim t^{2H-1}$, where $H>0$. The particles independently reset to the origin, and each particle's clock is set to zero upon spatial resetting.  Employing the known steady-state position distribution of a \emph{single} particle undergoing the sBm with renewal resetting [Bodrova et al., Phys. Rev. E \textbf{100}, 012120 (2019)], we study the  statistics of the system radius $\ell$ and of the center of mass (COM) of $N\gg 1$ particles. Typical fluctuations of $\ell$ fall under the Gumbel universality class for all $H>0$, and we use extreme value statistics to calculate the moments of $\ell$. We show that, for $H>1/2$, large deviations of the COM exhibit an anomalous scaling behavior. We also uncover a singularity in the corresponding rate function at $N\to\infty$, which is caused by a ``big jump" effect.

\end{abstract}




\end{frontmatter}


\section{Introduction}

Stochastic resetting represents a large class of stochastic processes where a random motion is instantaneously terminated at random times and restarted from a specified location. This paradigm is highly relevant in a broad range of natural processes and applications, see
reviews \cite{Evansreview,gupta2022stochastic,kundu2024preface}. 
Since Evans and Majumdar \cite{EM2011} formulated a simple general framework of stochastic resetting for diffusive processes, the topic has attracted considerable and sustained attention.
The problem becomes even richer and more interesting when there are many resetting agents, see Ref.~\cite{NagarGupta} for a review.

Stochastic resetting models differ among themselves by the nature of the random motion between the resetting events. Here we are interested in multi-particle systems such that, between the resetting events, the particles exhibit \emph{anomalous diffusion}, where the mean squared displacement of a particle scales with time as $t^{2H}$ \cite{metzler2000random, metzler2014anomalous, vilk2022unravelling}. The regime $0<H<1/2$ describes subdiffusion, while the regime $H>1/2$ corresponds to superdiffusion, which, for $H>1$, becomes super-ballistic. Standard diffusion is recovered at $H=1/2$. Subdiffusion ($H<1/2$) is observed in crowded cellular environments and viscoelastic media \cite{jeon2011vivo,metzler2000random}, superdiffusion ($1/2<H\leq 1$) characterizes active transport by molecular motors and L\'evy-walk-like animal foraging \cite{vilk2022unravelling,reynolds2015}, while super-ballistic transport ($H>1$) 
has been observed, for example, in commuting vultures which accelerate upon leaving the nest \cite{vilk2022unravelling}.

Anomalous transport with stochastic resetting has been studied for continuous\! -time random walks, scaled Brownian motion, and other anomalous diffusion processes \cite{maso2019transport, biswas2025resetting,Bodrova2019,Bodrova2019a,wang2022restoring,liang2025ultraslow, vinod2022time}. A simple way to model anomalous diffusion, that we adopt here, is scaled Brownian motion (sBm): a Gaussian process with independent increments, characterized by a power-law dependence of the diffusion coefficient on time, $D(t)\sim t^{2H-1}$, where $H$ can be any positive number \cite{Lim,Jeon,Safdari}. 

Stochastic resetting can be added to sBm in two different ways \cite{Bodrova2019,Bodrova2019a}. In the first one -- the nonrenewal resetting -- the spatial resetting events do not affect the time-dependence of the diffusion coefficient: the particle does not forget its local time \cite{Bodrova2019a}. In the second one -- the renewal resetting -- the local clock of the particle is reset to zero upon the particle's spatial resetting \cite{Bodrova2019}. The differences between these two settings were thoroughly discussed 
in Refs.~\cite{Bodrova2019,Bodrova2019a}. In particular, for the nonrenewal case the probability distribution of the particle position remains non-stationary at all times. In contrast, under renewal resetting the position distribution reaches a nonequilibrium steady state whose tails are stretched or compressed exponentials (see below). 
In this work we study the latter case and focus on the collective properties of $N\gg 1$ independent scaled Brownian particles subject to renewal resetting. Specifically, we study the statistics of the system radius and the statistics of the center of mass (COM).

A combination of anomalous diffusion and stochastic resetting of multiple agents arises naturally in several physical and biological contexts. A prominent example is central-place foraging, where independently foraging animals---such as bees, birds, or marine predators---perform anomalous, often superdiffusive, search movements and intermittently return to a central location such as a nest or hive \cite{reynolds2015,viswanathan1996,humphries2010,vilk2022ergodicity}. The return to the central place constitutes stochastic resetting with renewal, as each foraging bout effectively restarts from the home location. In this setting, the system radius characterizes the colony's spatial coverage, while the COM captures the overall displacement of the colony's centroid. Another relevant setting is intracellular transport, where cargo carried by molecular motors undergoes active transport with stochastic detachment events that effectively reset the cargo to a reference location \cite{metzler2014anomalous,jeon2011vivo}. 

Here are the main results of this work. Employing the known steady-state position distribution of a \emph{single} particle undergoing sBm with renewal resetting \cite{Bodrova2019}, we determine the full statistics of the system radius
$\ell$ (defined as the maximum distance of a particle from the origin), and of the COM. We show that the typical fluctuations of $\ell$ fall within the Gumbel universality class for all $H>0$. In contrast, the large deviations of the  COM behave very differently for $H<1/2$ and $H>1/2$. Remarkably, for $H>1/2$ the COM large-deviation statistics exhibit anomalous scaling behavior and a singularity (a jump in the first derivative) in the associated large deviation function  in the limit of $N\to \infty$. These features are caused by a ``big-jump'' mechanism (see e.g., Ref. \cite{Barkai2}), where a single particle wanders far from the rest of the particles and dominates the statistics. Finally, all our analytical results, obtained for the sBm under renewal resetting, also apply to the fractional Brownian motion \cite{mandelbrot1968fractional} under renewal resetting, as we explain in the following.

The remainder of the paper is structured as follows. In Sec. \ref{sec:microscopic_model} we recall the definition of the sBm. In Secs. \ref{sec:full_statistics} and \ref{sec:COM} we determine the complete statistics of the radius and of the COM of the system, respectively.  Finally, in Sec.~\ref{discussion} we summarize our results and discuss some promising directions for future work.  Throughout the paper we compare our results with Monte-Carlo simulations.

\section{Model} \label{sec:microscopic_model}

We consider \(N \gg 1\) independent particles moving on the real line. Each particle performs the sBm \cite{Lim,Jeon,Safdari} with  exponent $H>0$ and diffusion coefficient \(D\):
\begin{equation}
\label{micro}
\langle[x_i(t)-x_i(s)]^2\rangle = 2 D\,|t-s|^{2H},\qquad i=1,\dots,N.
\end{equation}
At random times one particle is randomly chosen with Poisson rate \(N r\) and reset to the origin. The particle's clock is also reset to zero, and the particle resumes the sBm.  The single-particle ($N=1$) variant of this model was introduced and analyzed by Bodrova et al. \cite{Bodrova2019}.

At times much longer than $1/r$ this ensemble of particles reaches a nonequilibrium steady state (NESS), and our objective is to characterize this NESS. To this end we will calculate the probability distributions of the system radius $\ell$ and of the system COM. Since the particles are non-interacting, these two distributions are completely determined by the steady-state position distribution $p_{\text{s}}(x)$ in the single-particle case, $N=1$. The latter distribution was calculated by Bodrova et al. \cite{Bodrova2019}, and it reads:
\begin{equation}\label{psA}
p_\text{s}(x) = \frac{1}{\sqrt{4\pi}}\int_{0}^{\infty} d\tau\,\frac{\,e^{-\frac{x^2}{4 \tau ^{2 H}}-\tau }}{\tau ^{H}}\,.
\end{equation}
Here, and quite often in the following, we set $r=D=1$. This corresponds to a proper choice of units of distance,  $\sqrt{D}/r^{H}$, and time, $1/r$.

The integral in Eq.~(\ref{psA})  does not allow a convenient exact evaluation. However, as we will see shortly, the most important properties of the two statistics, that we are going to focus on, are determined by the large-$|x|$ tail of $p_\text{s}(x)$. This tail can be determined  via a saddle-point evaluation of the integral in Eq.~(\ref{psA}), and one obtains \cite{Bodrova2019,MV2025}
\begin{equation}\label{largexA}
p_\text{s}(|x|\!\gg\! 1)\simeq  \left(\frac{H}{2}\right)^{\frac{1-2 H}{2 (2 H+1)}} \frac{| x| ^{\frac{1-2 H}{2 H+1}}}{\sqrt{2 (2 H+1)} } \exp\left[-\frac{(2 H+1) \left(\frac{H}{2}\right)^{\frac{1}{2 H+1}}| x| ^{\frac{2}{2 H+1}}}{2 H}\right].
\end{equation}
This tail is super-exponential for $H<1/2$ and sub-exponential for $H>1/2$.  In the marginal case of the  standard Brownian motion, $H=1/2$, the tail is exponential and coincides with the exact single-particle steady-state distribution $p_\text{s}(x)=(1/2) \,e^{-|x|}$ \cite{EM2011}. 

For reference purposes,  we also present the $|x|\!\ll\! \ell$ asymptotics of $p_\text{s}(x)$~\cite{MV2025}:
\begin{equation}
\label{smallxA}
p_\text{s}(|x| \!\ll \! \ell_0) \!\simeq\! \frac{1}{\sqrt{4\pi}}\times
\begin{cases}\Gamma(1\!-\!H) \!- \!\Gamma(1\!-\!3H)\frac{x^2}{4} ,&  \text{for $0<H<1/3$},\\ 
\Gamma(1\!-\!H)\!+\! 
\frac{1}{2H}\Gamma\left(\frac{1}{2}\!-\!\frac{1}{2H} \right) \left(\frac{x^2}{4}\right)^{\frac{1-H}{2H}},& \text{for $1/3<H<1$},\\
\frac{1}{2H}\Gamma\left(\frac{1}{2}\!-\!\frac{1}{2H}\right)\left(\frac{x^2}{4}\right)^{\frac{1-H}{2H}}, & \text{for $H>1$}\,,
\end{cases}
\end{equation}
where $\Gamma(z) = \int_0^{\infty} t^{z-1} e^{-t} dt$ is the gamma function. According to Eq.~(\ref{smallxA}), the maximum density is observed at $x=0$, and it is finite only for $H<1$. For $H\geq 1$ it diverges, but this singularity is integrable. 

Throughout this work we compare our analytical results with Monte Carlo simulations. The simulations are event-driven: resets are Poissonian at total rate $Nr$, so the time between consecutive events is exponentially distributed and serves as the time step $\Delta t$. Between resets, the coordinate of each particle $i$ is updated by a Gaussian increment
with variance $4H\,\tau_i^{2H-1}\Delta t$, where $\tau_i$ is the time since the last reset.
The system relaxes to its steady state on a timescale $\mathcal{O}(1/r)$, so long runs provide many effectively uncorrelated steady-state samples \cite{MV2025}. Simulation parameters for each figure are given in the corresponding captions.


\section{Full statistics of the system radius $\ell$}
\label{sec:full_statistics}
\begin{figure}[ht]
\centering
\includegraphics[clip,width=0.60\textwidth]{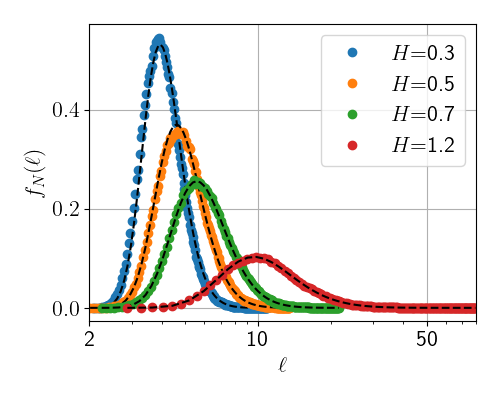}
\caption{Steady-state distribution of the system radius $\ell$ for different values of $H$ (see legend). Lines: Eq. \eqref{eq:f_N} , evaluated numerically. Symbols: simulations, where $N = 100$, and the simulation time is $t=10^5$.}
\label{fig:f_N_ell}
\end{figure}

Building on the single-particle NESS distribution \cite{Bodrova2019}, summarized in Sec.~\ref{sec:microscopic_model}, we now turn to collective statistics of the $N$-particle system, and we start with determining the full statistics of the system radius
$\ell$. To this end we integrate the  single-particle distribution ${p}_\text{s}(x)$ in Eq.~(\ref{psA}) over $x\in[-\ell,\ell]$ and arrive at the cumulative single-particle probability
\begin{equation}\label{eq:I_def}
I(\ell) =
\int_{0}^{\infty} e^{-\tau}\,
\mathrm{erf}\!\left(\frac{\ell}{2}\tau^{-H}\right)\, d\tau\,.
\end{equation}
Since the particles are independent, the cumulative probability that the system radius does not exceed $\ell$ can be written as
\begin{equation}\label{eq:F_N}
F_N(\ell) = [I(\ell)]^{N}.
\end{equation}
Differentiating this expression with respect to $\ell$, we obtain the exact probability density of the radius $\ell$:
\begin{equation}\label{eq:f_N}
f_N(\ell) = N\,I_\ell'(\ell)\,[I(\ell)]^{N-1},
\end{equation}
where, 
\begin{equation}
    I_\ell'(\ell) = \frac{1}{\sqrt{\pi}}\int_{0}^{\infty} \tau^{-H} \exp\!\left(-\tau - \frac{\ell^2}{4}\tau^{-2H}\right)d\tau.
\end{equation}
Fig. \ref{fig:f_N_ell} shows  numerically evaluated $f_N(\ell)$ alongside with our simulation results for different $H$.

The average radius $\bar{\ell}$ is given by the first moment of $f_N(\ell)$:
\begin{equation}\label{eq:mean_ell}
\bar{\ell}
= \int_{0}^{\infty} \ell\, f_N(\ell)\, d\ell
= \int_{0}^{\infty} [1 - F_N(\ell)]\, d\ell
= \int_{0}^{\infty} [1 - I(\ell)^{N}]\, d\ell, 
\end{equation}
where we have integrated by parts and used Eq. \eqref{eq:F_N}.  For the standard Brownian motion, $H=1/2$, Eq.~(\ref{eq:mean_ell}) yields $I(\ell) = 1 - e^{-\ell}$, and we obtain
\begin{equation}\label{eq:mean_Half}
\bar{\ell}= \int_{0}^{\infty} [1 - (1 - e^{-\ell})^{N}]\, d\ell
= H_N = \psi(N+1) + \gamma,
\end{equation}
in agreement with Ref. \cite{VAM2022}. Here $H_N$ is the $N$-th harmonic number, $\psi(x)=\Gamma'(x)/\Gamma(x)$ is the digamma function, and $\gamma=0.57721\dots$ is Euler's constant.

For general $H>0$ 
the integral in Eq.~\eqref{eq:mean_ell} is hard to evaluate analytically. Fortunately,  there is an important shortcut here, 
based on the observation that  the cumulative distribution $F_N(\ell)=I(\ell)^N$ is the distribution of the maximum of $N$ independent random numbers drawn from the same parent distribution $I(\ell)$. Therefore, we can apply to this case the well-developed theory of extreme–value statistics (EVS) of identically-distributed independent random numbers \cite{Fisher,Gnedenko,Gumbel,Leadbetter}, see Ref. \cite{Majumdar2020} for an accessible exposition. 
According to the EVS theory,  for $N\gg 1$, the average steady-state radius $\bar{\ell}$ is determined  by the \emph{tail} of the parent distribution $I(\ell)$. Once the tail is known, the EVS theory provides both the leading-order result for $\bar{\ell}$ and its universal subleading correction \cite{Majumdar2020}. Our task therefore reduces to obtaining the large--$\ell$ asymptotic behavior of the ``survival function"
\begin{equation}
\label{Il}
1-I(\ell)
= \int_{0}^{\infty} e^{-\tau}\, \mathrm{erfc}\!\left(\frac{\ell\,\tau^{-H}}{2}\right) d\tau .
\end{equation}
Using the identity
\[
\mathrm{erfc}(z)=\frac{2}{\sqrt{\pi}}\int_{z}^{\infty} e^{-s^{2}}\,ds,
\]
and noting that the integrand in Eq.~(\ref{Il}) is strictly positive,  we can exchange the order of integration (even though the inner integration limits depend on $\tau$). 
We obtain
\begin{eqnarray}
1-I(\ell)
&=&\frac{2}{\sqrt{\pi}}\int_{0}^{\infty}\!\int_{\frac{\ell}{2}\tau^{-H}}^{\infty}
e^{-\tau}e^{-s^{2}}\,ds\,d\tau \nonumber\\
&=&\frac{2}{\sqrt{\pi}}\int_{0}^{\infty} e^{-s^{2}}
\left[\int_{(\ell/2s)^{1/H}}^{\infty} e^{-\tau}d\tau\right] ds \nonumber \\
&=& \frac{2}{\sqrt{\pi}}\int_0^\infty e^{-\Psi(s,\ell)}ds \,, \label{oneminusI}
\end{eqnarray}
where
\[
\Psi(s,\ell)=s^{2}+\left(\frac{\ell}{2}\right)^{1/H}s^{-1/H}.
\]
Since we are interested in the  $\ell\gg 1$ asymptotic, we can evaluate the integral in Eq.~(\ref{oneminusI}) by the saddle-point method \cite{MathewsWalker}. Minimizing $\Psi(s,\ell)$ with respect to $s$ we obtain the saddle point
\begin{equation}\label{eq:saddle}
s_* = \left(\frac{1}{2H}\right)^{\!\frac{H}{2H+1}} \!\left(\frac{\ell}{2}\right)^{\!\frac{1}{2H+1}}.
\end{equation}
Evaluating $\Psi(s_*,\ell)$ gives the leading-order exponential term. 
To calculate the pre-exponential factor, we expand $\Psi(s,\ell)$  around $s=s_*$ up to the second order and perform Gaussian integration over $s$. Overall, the result is
\begin{equation}\label{eq:tail}
1-I(\ell)
\simeq
C_H\,\exp(-\kappa_H \ell^{\frac{2}{2H+1}})\,,
\end{equation}
where
\begin{equation}
\label{Candkappa}
C_H = \frac{2\sqrt{H}}{\sqrt{2H+1}}
\quad \text{and}\quad
\kappa_H = 2^{-\frac{2 (H+1)}{2 H+1}} H^{-\frac{2 H}{2 H+1}} (2 H+1)\,.
\end{equation}
As one can see from Eq.~(\ref{eq:tail}),  the $\ell \to \infty$ tail of the survival function $1-I(\ell)$ decays faster than a power law  for all $H>0$. Therefore, according to the EVS theory \cite{Majumdar2020}, the typical fluctuations of $\ell$ belong  to the Gumbel universality class. 
That is, after proper centering and rescaling,  the typical fluctuations are distributed, as $N\to \infty$,  as follows:
\[
\Pr\!\left(\frac{\ell - b_N}{a_N} \le z\right) \to e^{-e^{-z}},
\]
with the mean 
\begin{equation}\label{eq:bN}
b_N \simeq 
\kappa_H^{-\frac{2H+1}{2}} [\ln(C_H N)]^{\frac{2H+1}{2}}\,.
\end{equation}
The local scale of fluctuations around $b_N$ is determined by the inverse of the ``hazard rate" $h(\ell)=I'(\ell)/[1-I(\ell)]$, giving
\begin{equation}\label{eq:aN}
a_N \simeq \frac{2H+1}{2\,\kappa_H^{\frac{2H+1}{2}}} [\ln(C_H N)]^{H-1/2}.
\end{equation} 
Consequently, the average radius $\bar{\ell}$ has the following universal $N\to \infty$ asymptotic form:
\begin{equation}\label{eq:EL_final}
\bar{\ell} \simeq b_N + \gamma\,a_N 
= \kappa_H^{-\frac{2H+1}{2}}[\ln(C_H N)]^{H+1/2} + \gamma \frac{2H+1}{2\,\kappa_H^{\frac{2H+1}{2}}} [\ln(C_H N)]^{H-1/2}\,.
\end{equation}
Notably, the first term in Eq. \eqref{eq:EL_final} $b_n$ represents the most probable system radius, while the second term $\gamma a_N$ describes a universal correction originating from the Gumbel statistics.  Note that for $0<H<1/2$,  the second term decreases with an increase of $N$ so that the average radius is equal to $b_n$ without any correction which survives the limit of $N\to \infty$. In contrast, when $H>1/2$, the second term \emph{increases} with $N$, although slower than the first term.  For the marginal value $H=1/2$ the second term is constant $O(1)$, and we obtain $C_{1/2}=\kappa_{1/2}=1$,  and $\bar{\ell} \simeq \ln N +\gamma$, in agreement with Ref. \cite{VAM2022} and with the large-$N$ limit of Eq. \eqref{eq:mean_Half}.

\begin{figure}[t]
\centering
\includegraphics[clip,width=0.97\textwidth]{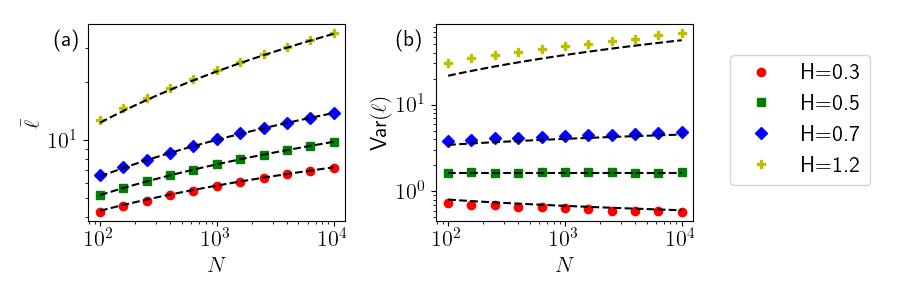}
\caption{(a) The average system radius $\bar{\ell}$, Eq. \eqref{eq:EL_final} and (b) the variance of the system radius, Eq. \eqref{eq:Var_asymp}, vs. $N$ for different values of $H$ (see legend). Lines: theoretical predictions. Symbols: simulation results, where the simulation time is $t=10^5$.} 
\label{fig:stat_ell}
\end{figure}

Using  the EVS,  we can also calculate higher moments of $\ell$. In particular, the variance is 
\begin{equation}
\mathrm{Var}(\ell)= \frac{\pi^2}{6}\,a_N^2 + o(a_N^2)\,\simeq \frac{\pi^2}{6} \frac{(2H+1)^2}{4\,\kappa_H^{2H+1}} [\ln(C_H N)]^{2H-1} \,.\label{eq:Var_asymp}
\end{equation}
For $H=1/2$ this gives $\mathrm{Var}(\ell)= \pi^2/6$, again in agreement with Ref. \cite{VAM2022}. In Fig. \ref{fig:stat_ell} we compare Eqs. \eqref{eq:EL_final} and \eqref{eq:Var_asymp} with our numerical simulations for different $H$. 

\section{Full statistics of the center of mass} \label{sec:COM}

Continuing to build on the single-particle statistics \cite{Bodrova2019}, we will now obtain our second main result: the large-deviation statistics of the COM. We consider the probability distribution $P(a,N)$ of observing the center of mass of the system in the steady state,
\begin{equation}\label{comcondition}
\bar{x} \equiv\frac{1}{N}\sum_{i=1}^{N} x_i\,,
\end{equation}
to be equal to a prescribed value $a$. Sometimes it will be more convenient to deal  with the probability distribution $\mathcal{P}(A,N)$  of the sum itself,
\begin{equation}
\Sigma\equiv \sum_{i=1}^{N} x_i =A\,.
\end{equation}
Since the particles are non-interacting, $\Sigma$ is a sum of $N\gg 1$ independent identically distributed random numbers. Similarly to the statistics of $\ell$, the statistics of $\Sigma$ is strongly affected by the behavior of the large-$|x|$ tail of the ``parent distribution" $p_s(x)$ \cite{Petrov,Nagaev,Denisov}. An inspection of the large-$|x|$ tail of the single-particle distribution~(\ref{largexA}) shows that there are two qualitatively different regimes of the scaling behavior of the probability distributions $P(a,N)$ and  $\mathcal{P}(\Sigma,N)$, depending on $H$. 

For $H\leq 1/2$ the large-$|x|$ tail of $p_s(x)$ falls off exponentially, or faster than exponentially, with $|x|$,  see Eq.~(\ref{largexA}). This regime corresponds to the ``standard" large-deviation scaling behavior of  $\mathcal{P}(A,N)$ at large $N$ \cite{Petrov}, 
\begin{equation}
\label{standard}
-\ln \mathcal{P}(A,N) \simeq  N f \left( \frac{A}{N} \right) \equiv N f \left(a\right)\,,
\end{equation}
where the rate function $f(a)$ remains analytic even in the limit of $N\to \infty$.

For $H>1/2$ the large-$|x|$ tail of $p_s(x)$ in Eq.~(\ref{largexA}) decays as a stretched exponential. This leads to an anomalous scaling behavior of the large deviations \cite{Barkai2,Petrov,Nagaev,Denisov}:
\begin{equation}
\label{anomalous}
-\ln \mathcal{P}(A,N) \simeq  N^{\mu} \phi \left( \frac{A}{N^{\nu}} \right)\,,
\end{equation}
where the anomalous exponents $\mu>0$ and $\nu>0$ are smaller than $1$. The underlying mechanism of the scaling anomaly is the ``big jump effect" observed at sufficiently large $A$, where a rare large deviation of the coordinate of a \emph{single} particle -- the big jump -- provides a dominant contribution to $A$ and determines its statistics. Remarkably, in the limit of  $N\to \infty$ while $A/N^{\nu} =\text{const}$, the rate function $\phi(y)$ exhibits a finite jump in its first derivative at some point $y_c$. Such a nonanalyticity can be interpreted as a first-order phase transition, see e.g. \cite{Smith2022}. Below the transition point $A=A_c$ all the particles give comparable contributions to $A$, and the fluctuations of $A$ are Gaussian, whereas above the transition point the rate function is affected by the big jump. An important attribute of this scenario is the regime of moderately large deviations at $A\gtrsim A_c$, where the two mechanisms - the Gaussian fluctuations and the big jump -- coexist, and the resulting rate function $\phi(\dots)$ is determined by their competition. 

\begin{figure}[t]
\centering
\includegraphics[clip,width=0.60\textwidth]{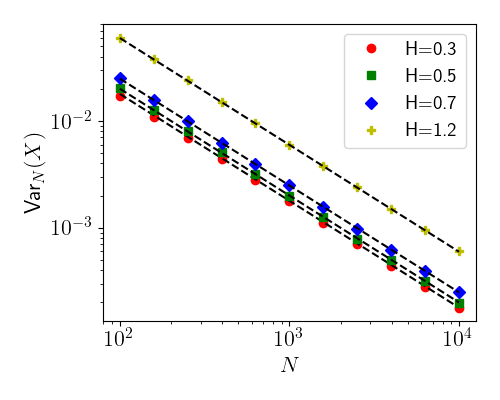}
\caption{Variance of the center of mass $X$ versus $N$, Eq. \eqref{varN}, for different values of $H$ (see legend). Lines: theoretical predictions. Symbols: simulation results, where the simulation time is $t=10^5$. }
\label{fig:stat_X}
\end{figure}

Below we consider the standard scaling regime $H\leq 1/2$ and the anomalous scaling regime $H>1/2$ separately. Prior to that, however, let us calculate the variance of the distributions $P(a,N)$ and $\mathcal{P}(A)$. The variance of the position of a \emph{single} particle is \cite{Bodrova2019}
\begin{equation}
  \sigma^2= \int_{-\infty}^{\infty} x^2 p_s(x)\,dx= 2 \Gamma (2 H+1)\,.
\end{equation}
By virtue of additivity of the variance of independent random variables, the variance of the center of mass $X$ is
\begin{equation}
\label{varN}
  \sigma^2_a =\frac{2 \Gamma (2 H+1)}{N}\,,
\end{equation}
which, in terms of the variance of the sum $\Sigma$, is equivalent to
\begin{equation}
\label{varNA}
  \sigma^2_A =\frac{N}{2\beta}\,, \quad \text{where} \quad \beta = \frac{1}{4 \Gamma(2H+1)}\,.
\end{equation}
Equations~(\ref{varN}) and (\ref{varNA})  hold for all $H>0$.  Fig. \ref{fig:stat_X} shows a comparison of Eq. \eqref{varN} with our numerical simulations.

\subsection{$H\leq 1/2$: standard scaling}

\begin{figure}[ht]
\centering
\includegraphics[clip,width=0.60\textwidth]{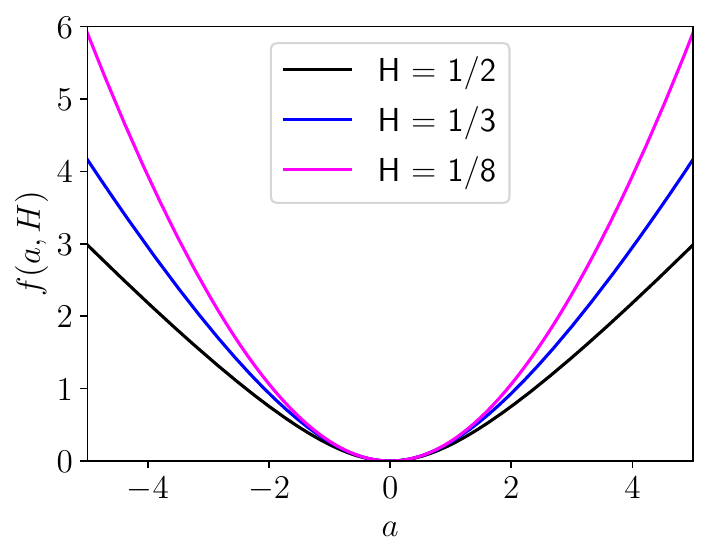}
\caption{The rate function $f(a;H)$, see Eqs.~(\ref{saddlepointeq2}) and (\ref{ratef}), which describes the full statistics of the center of mass of the system for $H=1/2$, $1/3$, and $1/8$, see legend.}
\label{ratefun}
\end{figure}

In this case the rate function $f(a)$ is given by Cramér's theorem, see e.g. Ref. \cite{Petrov}. Still, for completeness, we will calculate $f(a)$ from first principles. 
The starting point of this calculation is the exact one-particle position distribution $p_{\text{s}}(x)$ \cite{Bodrova2019}, described by Eq. (\ref{psA}).
The joint steady-state distribution $P(x_1, x_2, \dots, x_N)$ of the positions of all particles can be obtained from the single-particle distribution:
\begin{equation}\label{joint}
P(x_1, x_2, \dots, x_N)=\left(\frac{r}{\sqrt{4\pi}}\right)^N \prod_{i=1}^N \int_0^{\infty} e^{-\frac{x_i^2}{4 \tau^{2H}}-\tau} \tau^{-H}\,d\tau\,.
\end{equation}
Then the distribution $P(a,N)$ can be written as
\begin{equation}
P(a,N)  = \int_{-\infty}^{\infty} dx_1 \int_{-\infty}^{\infty} dx_2\,
\dots \int_{-\infty}^{\infty} dx_N \prod_{i=1}^{N} p_{\text{s}}(x_i)\,\delta \left(\frac{1}{N}\sum_{i=1}^N x_i -a\right). \label{proba}
\end{equation}
Using the exponential representation of the delta-function, we can rewrite this expression as
\begin{equation}
  P(a,N)  = \frac{N}{2\pi}\, \int_{-\infty}^\infty dk \,e^{-ikNa}\, \left[\int_{-\infty}^{\infty} dx\,e^{ikx} p_{\text{s}}(x)\right]^N.\label{proba1}
\end{equation}
The internal integral can be recast as follows:  
\begin{equation}\label{Ink1}
F(k)\equiv\int_{-\infty}^{\infty} dx\, e^{ikx} p_{\text{s}}(x)= \int_0^{\infty} d\tau\,\tau^{-H} e^{-\tau}
\int_{-\infty}^{\infty} dx \,e^{ikx-\frac{x^2}{4\tau^{2H}}} \,.
\end{equation}
The integration over $x$ is elementary, and we obtain
\begin{equation}\label{Ink2}
F(k) = \int_0^{\infty} d\tau\,e^{-\tau-k^2 \tau^{2H}}\,,
\end{equation}
which leads to the exact expression
\begin{equation}\label{proba3}
  P(a,N) = \frac{N}{2\pi} \int_{-\infty}^{\infty} dk\,e^{-i k N a+N \ln F(k)}\,.
\end{equation}
When $N\gg 1$ and $H<1/2$ we can evaluate the integral in Eq.~(\ref{proba3}) by the saddle-point method in the complex plane \cite{MathewsWalker}. Here it suffices for our purposes to find only the leading-order exponential dependence of $P(a,N)$. The saddle-point equation is
\begin{equation}
\label{saddlepointeq1}
\frac{F'(k)}{F(k)}=i a\,.
\end{equation}
For $H<1/2$ the saddle point $k_*$ is purely imaginary: $k_*=i \kappa$, where $\kappa$ is real, and Eq.~(\ref{saddlepointeq1}) takes the form
\begin{equation}
\label{saddlepointeq2}
a=2\kappa \,\frac{\int_0^{\infty} d\tau \,\tau^{2H}\,e^{-\tau+\kappa^2 \tau^{2H}}}{\int_0^{\infty} d\tau\,e^{-\tau+\kappa^2\tau^{2H}}}\,.
\end{equation}
At a given $a$, Eq.~(\ref{saddlepointeq2}) is an algebraic equation for $\kappa$. We are interested in the rate function $f(a,H)$, which determines $\mathcal{P}(a,N;H)$, up to subleading pre-exponential factors,  via the relation $-\ln P(a,N;H) \simeq N f(a,H)$, see Eq.~(\ref{standard}). The saddle-point evaluation yields
\begin{equation}
\label{ratef}
    f(a;H)=\kappa_*(a) a- \ln F[k_*(a)]\,.
\end{equation}
This expression reproduces Cramér's theorem,  see e.g. Ref. \cite{Petrov}. Equations~(\ref{saddlepointeq2}) and (\ref{ratef}) determine the rate function $f(a;H)$  in  a parametric form, where $\kappa$ plays the role of the parameter. Figure \ref{ratefun} shows plots of this rate function for $H=1/8$, $1/3$ and $1/2$. 

For the standard Brownian motion, $H=1/2$,  Eq.~(\ref{Ink2}) simplifies to $F(k)=(1+k^2)^{-1}$, and the relevant saddle point is
\begin{equation}\label{saddlepoint}
k=k_*(a)= i \left(\frac{1-\sqrt{a^2+1}}{a}\right)\,.
\end{equation}
This yields the rate function in a simple analytic form,
\begin{equation}\label{LDF}
f(a,1/2)  = \sqrt{a^2+1}+\ln \frac{2
   \left(\sqrt{a^2+1}-1\right)}{a^2}-1\,,
\end{equation}
in agreement with Ref. \cite{VAM2022}. In Fig. \ref{ratefun}  the plot of $f(1,1/2)$ is shown by the black curve. 

Back to general $H>0$, the small-$a$ region corresponds to the Gaussian fluctuations, which are dominated by the small-$k$ region of the integral over $k$ in Eq.~(\ref{proba3}). Here we can, \emph{for any} $H>0$, Taylor-expand the exponent $\exp(-Dk^2\tau^{2H})$ in the integrand. The final result, 
\begin{equation}
f(a\to 0,H) = \frac{a^2}{4 \Gamma(2H+1)}\,,
\end{equation}
agrees with the expression ~(\ref{varN}) for the variance, as to be expected.

\subsection{$H> 1/2$:  anomalous scaling and phase transition}

The contributions to the rate function $\phi(y)$ from the big jump,  $O(A^{\frac{2}{2H+1}})$ [which is determined by the 
$|x|\gg 1$ tail of the single-particle distribution, see Eq. (\ref{largexA})],  and from the Gaussian fluctuations, $O(A^2/N)$,  are comparable for $A \sim N^{\frac{2H+1}{4H}}$. This simple estimate correctly predicts the anomalous scaling behavior (\ref{anomalous}) with the exponents
\begin{equation}
\label{munu}
    \mu = \frac{1}{2H}\quad \text{and}\quad \nu = \frac{2H+1}{4H}\,,
\end{equation}
both of which are smaller than $1$ once $H>1/2$. The resulting rate function $\phi(y)$ -- which describes the coexistence of a big jump of a single particle and typical contributions from the rest of particles -- has the following form \cite{Nagaev}:
\begin{equation}
    \phi(y) = \inf_{0\leq z\leq y} \left[b(H) \,z^\frac{2}{2H+1}+\beta(H)(y-z)^2\right]\,,
\end{equation}
where the coefficients
\begin{equation}
 b(H)= \frac{2 H+1}{2 H} \left(\frac{H}{2 }\right)^{\frac{1}{2 H+1}} \quad \text{and}  \quad   \beta(H) = \frac{1}{4 \Gamma(2H+1)}
\end{equation}
follow from Eqs.~(\ref{smallxA}) and ~(\ref{varNA}), respectively.  The first-order transition occurs at the point
\begin{equation}
\label{yc}
    y=y_c(H) =\frac{2H}{2H-1} \left[\frac{(2H-1) b(H)}{(2H+1) \beta(H)}\right]^\frac{2H+1}{4H} \,. 
\end{equation}
At $y<y_c$ one obtains the Gaussian asymptotic $\phi(y)=\beta(H) y^2$, while at $y\gg y_c$, $\phi(y)$ asymptotically approaches $b(H) \,y^\frac{2}{2H+1}$.

\begin{figure}[t]
\centering
\includegraphics[clip,width=0.97\textwidth]{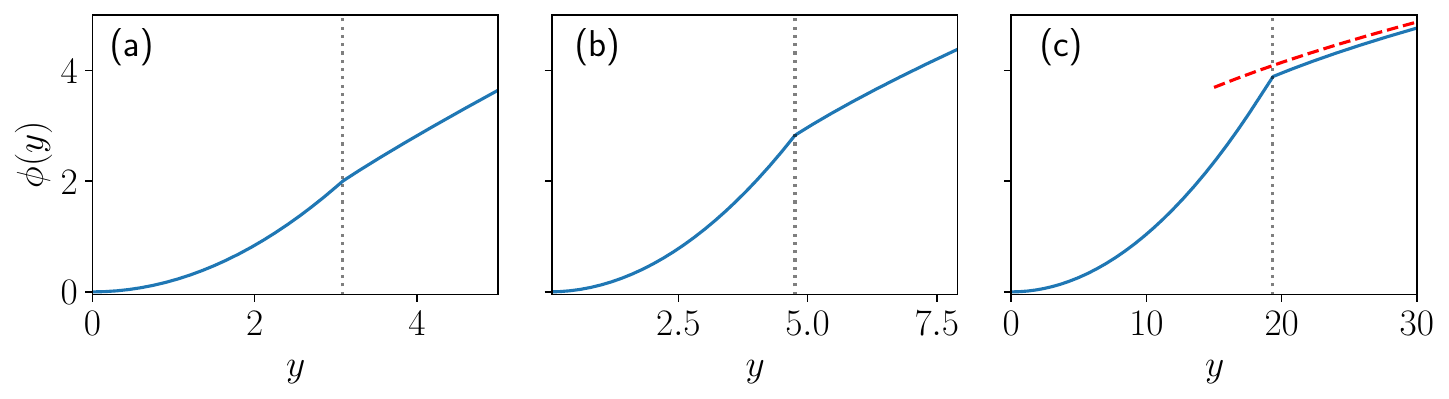}
\caption{The anomalous rate function $\phi(y)$, see Eqs.~(\ref{anomalous}) and~(\ref{munu}), for $H=2/3$ (a), $H=1$ (b), and $H=2$ (c). The critical point is described by Eq.~(\ref{yc}) and is marked in all panels by a dotted line. The red dashed line in panel c shows the large-$y$ asymptotic  $b \,y^\frac{2}{2H+1}$.}
\label{anomrate}
\end{figure}

Figure \ref{anomrate} shows plots of the rate function $\phi(y)$ for three different values of $H>1/2$ and the predicted critical point $y_c$ in each case.  One can see the two distinct behaviors of the rate function for $y<y_c(H)$ and $y>y_c(H)$. 

Table~\ref{tab:summary} summarizes the main features of the large-deviation statistics of the COM across the different regimes of $H$.

\begin{table}[ht]
\centering
\small
\begin{tabular}{@{}lccc@{}}
\hline\hline
 & $H<1/2$ & $H=1/2$ & $H>1/2$ \\
\hline
Tail of $p_s$ & super-exponential & exponential & stretched exponential \\
Large-deviation scaling & standard & standard & anomalous \\
Rate function & analytic & analytic & singular at $y_c$ \\
Phase transition & none & none & first order \\
\hline\hline
\end{tabular}
\caption{COM statistics for different $H$. See text for explicit expressions.}
\label{tab:summary}
\end{table}

\section{Discussion}
\label{discussion}

In this work we have analyzed collective statistical properties of a system of many scaled Brownian particles, independently resetting with renewal, by focusing on the statistics of the system radius and of the center of mass (COM).

Our first main result concerns the statistics of the system radius at the steady state. Since the steady-state position distribution of a single particle \cite{Bodrova2019} decays faster than a power law for all $H > 0$, the typical fluctuations that determine the moments of the distribution fall within the Gumbel universality class. 

Our second main result deals with the COM statistics. As we have seen, these statistics exhibit a qualitative change of behavior at $H = 1/2$. For $H < 1/2$, the system is in a ``condensed" state, which manifests itself in the conventional large-deviation statistics of the COM. In contrast, for $H > 1/2$ the system becomes susceptible to ``big jumps" where a single particle can wander anomalously far from the other particles and dominate the COM. This remarkable and quite universal, see e.g. Ref. \cite{Barkai2}, effect leads to an anomalous scaling of large deviations of the COM and, in the limit of $N\to \infty$, to a nonanalyticity in the associated rate function. This nonanalyticity can be classified as a first-order phase transition. The transition between the two regimes occurs at $H=1/2$.

Notably, these results, although obtained for the sBm,  also extend for the fractional Brownian motion \cite{mandelbrot1968fractional} under renewal resetting. This is because the single-particle NESS of the latter coincides with the one for the sBm \cite{wang2021timeaveraging}. Another possible extension of this theory is to heterogeneous diffusion processes (HDPs) \cite{wang2021timeaveraging,wang2022restoring}, where the diffusion coefficient depends on the position rather than on time.  In this case the single-particle NESS distribution is different, and all the analysis needs to be adapted. Still another  extension can address biological systems where the anomalous diffusion exponent $H$  varies across individuals, leading to a heterogeneous ensemble where the particles are no longer identically distributed.

Importantly, the collective statistical properties of the scaled Brownian particles, that we studied here, are ultimately determined by the properties of the single-particle steady state. This simplicity, however, is special to \emph{noninteracting} particles. In many physical and biological systems anomalously diffusing particles experience strong interactions, crowding, or coupling through nonlocal reset rules, and their collective fluctuations cannot be readily inferred from single-particle behavior alone~\cite{NagarGupta,VAM2022,Berestycki1,Berestycki2,SSM,vatash2025many,biroli2023extreme,biroli2025resetting}. A first step in developing an adequate theoretical framework for interacting ensembles of anomalously diffusing particles under resetting has been recently made in Ref. \cite{MV2025} in hydrodynamic limit $N\to \infty$. The next step could be developing
a fluctuating hydrodynamic theory that would be able to capture macroscopic fluctuations of these systems, such as the radius and the COM. For the standard Brownian motion of globally interacting particles with different resetting rules, a fluctuating hydrodynamics was developed in Refs. \cite{VAM2022,SSM}. Its extension to anomalous diffusion such as the scaled Brownian motion therefore presents an interesting  direction for future work.

\bigskip
\noindent
{\bf Acknowledgments}. We are grateful to Naftali R. Smith for valuable comments. B. M.  was supported by the Israel Science Foundation 
(Grant No. 1579/25).

 \bibliographystyle{elsarticle-num} 
 \bibliography{cas-refs}



\end{document}